\documentclass[ALICE,manyauthors]{cernphprep}

\newcommand{\hyp}{\ensuremath{_{\Lambda}^{3}\mathrm{H}}\xspace}
\newcommand{\antihyp} {\ensuremath{^3_ {\overline{\Lambda}}\overline{\mathrm{H}}}\xspace}
\newcommand{\he}   {\ensuremath{^3\mathrm{He}}\xspace}

\newcommand{\dectwo}{\ensuremath{\hyp \rightarrow {}^{3}\mathrm{He} + \pim}\xspace}

\usepackage[comma,square,numbers,sort&compress]{natbib}
\usepackage{hyperref}
\usepackage{lineno}
\usepackage{xspace}
\usepackage{booktabs} 
\usepackage{multirow}
\usepackage{bm}
\usepackage[T1]{fontenc}
\usepackage{orcidlink}

\begin{document}
%

\newcommand{\pp}           {pp\xspace}
\newcommand{\ppbar}        {\mbox{$\mathrm {p\overline{p}}$}\xspace}
\newcommand{\XeXe}         {\mbox{Xe--Xe}\xspace}
\newcommand{\PbPb}         {\mbox{Pb--Pb}\xspace}
\newcommand{\pA}           {\mbox{pA}\xspace}
\newcommand{\pPb}          {\mbox{p--Pb}\xspace}
\newcommand{\AuAu}         {\mbox{Au--Au}\xspace}
\newcommand{\dAu}          {\mbox{d--Au}\xspace}

\newcommand{\s}            {\ensuremath{\sqrt{s}}\xspace}
\newcommand{\snn}          {\ensuremath{\sqrt{s_{\mathrm{NN}}}}\xspace}
\newcommand{\pt}           {\ensuremath{p_{\rm T}}\xspace}
\newcommand{\meanpt}       {$\langle p_{\mathrm{T}}\rangle$\xspace}
\newcommand{\ycms}         {\ensuremath{y_{\rm CMS}}\xspace}
\newcommand{\ylab}         {\ensuremath{y_{\rm lab}}\xspace}
\newcommand{\etarange}[1]  {\mbox{$\left | \eta \right |~<~#1$}}
\newcommand{\yrange}[1]    {\mbox{$\left | y \right |~<~#1$}}
\newcommand{\dndy}         {\ensuremath{\mathrm{d}N_\mathrm{ch}/\mathrm{d}y}\xspace}
\newcommand{\dndeta}       {\ensuremath{\mathrm{d}N_\mathrm{ch}/\mathrm{d}\eta}\xspace}
\newcommand{\avdndeta}     {\ensuremath{\langle\dndeta\rangle}\xspace}
\newcommand{\dNdy}         {\ensuremath{\mathrm{d}N_\mathrm{ch}/\mathrm{d}y}\xspace}
\newcommand{\Npart}        {\ensuremath{N_\mathrm{part}}\xspace}
\newcommand{\Ncoll}        {\ensuremath{N_\mathrm{coll}}\xspace}
\newcommand{\dEdx}         {\ensuremath{\textrm{d}E/\textrm{d}x}\xspace}
\newcommand{\RpPb}         {\ensuremath{R_{\rm pPb}}\xspace}

\newcommand{\nineH}        {$\sqrt{s}~=~0.9$~Te\kern-.1emV\xspace}
\newcommand{\seven}        {$\sqrt{s}~=~7$~Te\kern-.1emV\xspace}
\newcommand{\twoH}         {$\sqrt{s}~=~0.2$~Te\kern-.1emV\xspace}
\newcommand{\twosevensix}  {$\sqrt{s}~=~2.76$~Te\kern-.1emV\xspace}
\newcommand{\five}         {$\sqrt{s}~=~5.02$~Te\kern-.1emV\xspace}
\newcommand{\twosevensixnn}{$\sqrt{s_{\mathrm{NN}}}~=~2.76$~Te\kern-.1emV\xspace}
\newcommand{\fivenn}       {$\sqrt{s_{\mathrm{NN}}}~=~5.02$~Te\kern-.1emV\xspace}
\newcommand{\LT}           {L{\'e}vy-Tsallis\xspace}
\newcommand{\GeVc}         {$\mathrm{Ge\kern-.1emV/}c$\xspace}
\newcommand{\MeVc}         {Me\kern-.1emV/$c$\xspace}
\newcommand{\TeV}          {Te\kern-.1emV\xspace}
\newcommand{\GeV}          {Ge\kern-.1emV\xspace}
\newcommand{\MeV}          {Me\kern-.1emV\xspace}
\newcommand{\GeVmass}      {Ge\kern-.2emV/$c^2$\xspace}
\newcommand{\MeVmass}      {Me\kern-.2emV/$c^2$\xspace}
\newcommand{\lumi}         {\ensuremath{\mathcal{L}}\xspace}

\newcommand{\ITS}          {\rm{ITS}\xspace}
\newcommand{\TOF}          {\rm{TOF}\xspace}
\newcommand{\ZDC}          {\rm{ZDC}\xspace}
\newcommand{\ZDCs}         {\rm{ZDCs}\xspace}
\newcommand{\ZNA}          {\rm{ZNA}\xspace}
\newcommand{\ZNC}          {\rm{ZNC}\xspace}
\newcommand{\SPD}          {\rm{SPD}\xspace}
\newcommand{\SDD}          {\rm{SDD}\xspace}
\newcommand{\SSD}          {\rm{SSD}\xspace}
\newcommand{\TPC}          {\rm{TPC}\xspace}
\newcommand{\TRD}          {\rm{TRD}\xspace}
\newcommand{\VZERO}        {\rm{V0}\xspace}
\newcommand{\VZEROA}       {\rm{V0A}\xspace}
\newcommand{\VZEROC}       {\rm{V0C}\xspace}
\newcommand{\Vdecay} 	   {\ensuremath{V^{0}}\xspace}

\newcommand{\ee}           {\ensuremath{e^{+}e^{-}}} 
\newcommand{\pip}          {\ensuremath{\pi^{+}}\xspace}
\newcommand{\pim}          {\ensuremath{\pi^{-}}\xspace}
\newcommand{\kap}          {\ensuremath{\rm{K}^{+}}\xspace}
\newcommand{\kam}          {\ensuremath{\rm{K}^{-}}\xspace}
\newcommand{\pbar}         {\ensuremath{\rm\overline{p}}\xspace}
\newcommand{\kzero}        {\ensuremath{{\rm K}^{0}_{\rm{S}}}\xspace}
\newcommand{\lmb}          {\ensuremath{\Lambda}\xspace}
\newcommand{\almb}         {\ensuremath{\overline{\Lambda}}\xspace}
\newcommand{\Om}           {\ensuremath{\Omega^-}\xspace}
\newcommand{\Mo}           {\ensuremath{\overline{\Omega}^+}\xspace}
\newcommand{\X}            {\ensuremath{\Xi^-}\xspace}
\newcommand{\Ix}           {\ensuremath{\overline{\Xi}^+}\xspace}
\newcommand{\Xis}          {\ensuremath{\Xi^{\pm}}\xspace}
\newcommand{\Oms}          {\ensuremath{\Omega^{\pm}}\xspace}
\newcommand{\degree}       {\ensuremath{^{\rm o}}\xspace}

\begin{titlepage}
\PHyear{2024}       
\PHnumber{148}      
\PHdate{27 May}     

\title{Measurement of ${}_{\Lambda}^{3}\mathrm{H}$\xspace production in Pb--Pb collisions at $\sqrt{s_{\mathrm{NN}}}$ = 5.02~Te\kern-.1emV\xspace}
\ShortTitle{Measurement of ${}_{\Lambda}^{3}\mathrm{H}$\xspace production in Pb--Pb collisions at \fivenn}   

\Collaboration{ALICE Collaboration\thanks{See Appendix~\ref{app:collab} for the list of collaboration members}}
\ShortAuthor{ALICE Collaboration} 

\begin{abstract}
The first measurement of \hyp and \antihyp differential production with respect to transverse momentum and centrality in \PbPb collisions at \fivenn is presented. The \hyp has been reconstructed via its two-charged-body decay channel, i.e., \dectwo. A Blast-Wave model fit of the \pt-differential spectra of all nuclear species measured by the ALICE collaboration suggests that the \hyp kinetic freeze-out surface is consistent with that of other nuclei. The ratio between the integrated yields of \hyp and \he is compared to predictions from the statistical hadronisation model and the coalescence model, with the latter being favoured by the presented measurements.

\end{abstract}
\end{titlepage}

\setcounter{page}{2} 

\section{Introduction} 

Relativistic heavy-ion collisions provide a rich environment for the study of nuclei, hypernuclei, and their charge conjugates. 
Hypernuclei are nuclei that have at least one hyperon among their constituents. The additional strange baryon leads to a unique nuclear system that can help to benchmark the interactions among hyperons and nucleons. 
Understanding such interactions has become particularly relevant in recent years due to their important connections with the modelling of the equation-of-state of dense astrophysical objects, such as the neutron stars~\cite{Schaffner-Bielich:2008zws, Tolos:2020aln}. 
In fact, a precise knowledge of the two-body (Y--N) and three-body (Y--N--N) hyperon--nucleon forces allows for determining whether the presence of hyperons in the innermost core of neutron stars, which is known to be energetically favourable, reconciles with the recent observations of large-mass neutron stars~\cite{Lonardoni:2014bwa,Logoteta:2019utx}. 
Multiple studies analysing particle momentum correlations~\cite{Fabbietti:2020bfg,ALICE:2020mfd} play a direct role in establishing these interactions. In a complementary approach, the lifetime and the binding energy of a hypernucleus reflect the strength of the Y--N and Y--N--N forces~\cite{Kamada:1997rv, Haidenbauer:2019boi, Lonardoni:2017uuu}.   

The lightest known hypernucleus is the hypertriton (\hyp), composed of a proton, a neutron, and a $\Lambda$ hyperon. As such, it represents the simplest system to study the interaction among $\Lambda$ hyperons and nucleons.
Early experiments studied \hyp using nuclear emulsions and helium bubble chambers~\cite{prem_lifetimes_1964, Keyes:1968zz, Phillips:1969uy, Bohm:1970se, Keyes:1973two}. These experiments obtained results with large statistical uncertainties. In recent years, more precise measurements have been performed by ALICE~\cite{ALICE:2015oer, ALICE:2019vlx, ALICE:2021sdc, ALICE:2022sco} and STAR~\cite{STAR:2017gxa, STAR:2021orx, STAR:2023fbc, Chen:2023mel, ma_hypernuclei_2023} Collaborations in relativistic heavy-ion collisions at the CERN LHC and BNL RHIC colliders. 
Such measurements contributed in determining that the \hyp is a loosely bound state with a $\Lambda$ separation energy, $B_\Lambda$, of the order of one hundred keV, and has an average lifetime very close to the free $\Lambda$ one~\cite{HypernuclearDataBase}. With such properties, the hypertriton wave function is expected to have a radial extension of approximately 5 fm~\cite{Cobis_1997, Nemura:1999qp, Hildenbrand:2019sgp}, considerably larger than the other nuclei with a mass number equal to three and comparable in size to the fireball created in \PbPb collisions.

There are two main classes of models used to explain the formation of nuclei in Pb--Pb collisions: the statistical hadronisation model (SHM) and the coalescence model. While the SHM is insensitive to the structure of the particle being produced~\cite{Andronic:2017pug, Vovchenko:2018fiy, Vovchenko:2019kes, Vovchenko:2022xil, Sharma:2022poi}, the coalescence model relies on the knowledge of the particle wave function to produce predictions~\cite{Chen:2003ava, Blum:2017qnn, Bellini:2018epz, Sun:2018mqq, Bellini:2020cbj, Kachelriess:2020amp,Mahlein:2023fmx}. There is a wealth of variations for each class of models but based on similar approach.
In the SHM, the production cross section of a specific particle is determined solely by a temperature, denoted as chemical freeze-out temperature, the volume of the system created in the collision, the quantum numbers, and the mass of the particle. In Pb--Pb collisions, the grand canonical formulation of the SHM has proven to be very successful in describing the yields of light flavoured hadrons~\cite{Andronic:2017pug}. Based on the two parameters from the analysis of light hadrons, the SHM then provides a parameter-free  prediction of the yields of (hyper)nuclei.
In the coalescence model, the production rate of nuclei is given by the convolution between the phase space of the nucleons produced in the collision and the nuclear wave function, according to the Wigner function formalism~\cite{Scheibl:1998tk}.
These two models give similar predictions for the yield ratios of most ordinary nuclei (e.g.~d/p)~\cite{Sun:2018mqq, Bellini:2018epz, Bellini:2020cbj} and qualitatively describe the available experimental data at the LHC~\cite{ALICE:2015wav,ALICE:2017xrp,ALICE:2017jmf,ALICE:2017nuf,ALICE:2019dgz,ALICE:2019fee,ALICE:2020foi,ALICE:2022pbb,ALICE:2021mfm,ALICE:2021ovi,ALICE:2022veq,ALICE:2023qyl}. 
However, when it comes to a tight configuration or a loosely-bound state, the prediction of the SHM and the coalescence model for particle ratios diverge significantly in all collision systems. 
For example, recent studies on the yield of {\ensuremath{^4\mathrm{He}}\xspace} in Pb--Pb collisions show an agreement with the SHM while the coalescence model failed to explain it~\cite{ALICE:2023qyl}. Conversely, the \hyp/$\Lambda$ ratio in p--Pb collisions~\cite{ALICE:2021sdc} is only well-described by the coalescence predictions.
This discrepancy indicates that the nucleosynthesis mechanism~\cite{Mrowczynski:2020ugu, Rais:2022gfg} still needs further investigation with new experimental measurements.

This letter presents the first measurement of the \hyp production in \PbPb collisions at a center-of-mass energy per nucleon pair \fivenn. In Sec.~\ref{sec:analysis} the ALICE apparatus and the analysis method are described, whereas the results are discussed in Sec.~\ref{sec:results}.

\section{The ALICE detector and analysis details\label{sec:analysis}}

The ALICE apparatus is composed of several specialised detectors, as detailed in~\cite{ALICE:2014sbx}. In this analysis, \hyp are reconstructed using both the Inner Tracking System (ITS)~\cite{ALICE:2010tia} and the Time Projection Chamber (TPC)~\cite{ALME2010316} detectors. The ITS is the closest central-barrel detector to the beam pipe and comprises six cylindrical layers. 
The primary role of ITS is to reconstruct charged-particle trajectories and to measure precisely the position of the interaction vertices. The TPC is a gaseous detector employed as the primary tracking and particle identification (PID) device in ALICE. The TPC is used to identify charged particles by measuring their specific energy loss.
Both ITS and TPC are placed in a homogeneous magnetic field of 0.5 T produced by a solenoidal magnet. These two detectors cover the entire azimuth in the pseudorapidity range $|\eta| <$ 0.9.
The centrality determination and trigger are provided by a pair of forward (2.8 $< \eta <$ 5.1) and backward ($-$3.7 $< \eta <$ $-$1.7) detectors called V0A and V0C~\cite{ALICE:2013axi}, respectively. The coincidence of a signal in the V0A and a signal in the V0C is used as a minimum-bias trigger. Additionally, two thresholds on the minimum amount of charge deposited on the V0 detector are implemented to trigger online on central and semi-central \PbPb collisions. These thresholds are defined by the Glauber model fit to the V0 detector signal amplitudes~\cite{ALICE:2013axi, ALICE:2018tvk}.

The data sample employed in this analysis has been collected during the LHC 2018 \PbPb run at \fivenn. 
Based on the signal amplitudes of the V0 detectors, three centrality classes are considered (from central to peripheral collisions): 0--10\%, 10--30\%, and 30--50\%. The centrality classes are labelled according to the corresponding percentiles of the Pb--Pb hadronic interaction cross section. The stored events, categorized into the 0--10\% and 30--50\% classes, are enriched by the online trigger on central and semi-central collisions.
In total, approximately 210 million events are studied. At the LHC energies, approximately the same number of \hyp and \antihyp are produced~\cite{ALICE:2023ulv}. 
Therefore, the results of \hyp and \antihyp are averaged. In the following, we use the notation \hyp for both particle and antiparticle.

A dedicated Monte Carlo (MC) simulated sample has been employed for optimising the \hyp selection and evaluating the efficiency corrections. The MC sample consists of \hyp signal injected on top of underlying \PbPb collisions, which are simulated with the \textsc{HIJING} event generator~\cite{Wang:1991hta}. In each MC event, 80 \hyp are injected. The transverse momentum (\pt) distribution of the injected signal is given by the Blast-Wave~\cite{Schnedermann:1993ws} function, with parameters taken from fit to the \he \pt spectra~\cite{ALICE:2022veq}. The particle transport through the detector material is done using \textsc{Geant4}~\cite{Agostinelli:2002hh}, which simulates both the interaction with the material and the weak decay kinematics of the \hyp.

In this analysis, \hyp candidates are reconstructed via the two-charged-body mesonic decay channel \mbox{\dectwo} (and the related
charge conjugated process).
Firstly, a preselection method is performed.
Tracks that have specific energy loss in the TPC compatible within 5$\sigma$ to that of \he or $\pi$ tracks are employed to reconstruct the \hyp decay topology with the algorithm used in previous analyses~\cite{ALICE:2015oer, ALICE:2019vlx, ALICE:2021sdc}.
The reconstruction of \hyp candidates employed the ALICE secondary vertex finder method which pairs opposite-sign \he and $\pi$ candidate tracks. 
The tracks are required to have more than 50 hits in TPC ($N^{\mathrm{TPC}}_{\mathrm{cls}}$) and good fit quality ($\chi^2/N^{\mathrm{TPC}}_{\mathrm{cls}} < 4 $).
In order to mitigate the contamination from \he produced in the interaction with materials, an additional selection $\pt > 1.2$ \GeVc is imposed for \he.
Additional selection criteria are used by combining data on the decay kinematics and the position of the decay vertex. The selections include the maximum distance-of-closest-approach (DCA) of the decay products to the primary vertex ($\mathrm{DCA_{toPV}}$ $<$ 8 cm) and among the daughter tracks themselves ($\mathrm{DCA_{tracks}}$ $<$ 1.6 cm), and the cosine of the angle between the \hyp momentum vector and the straight line connecting the primary and secondary vertices ($\cos\mathrm{\theta_P}$). The selection on $\cos\mathrm{\theta_P}$ is set as a function of the proper decay time of the candidates such to ensure a 95\% efficiency for reconstructing true \hyp.

\begin{figure}[ht]
    \centering
    \includegraphics[width=0.6\textwidth]{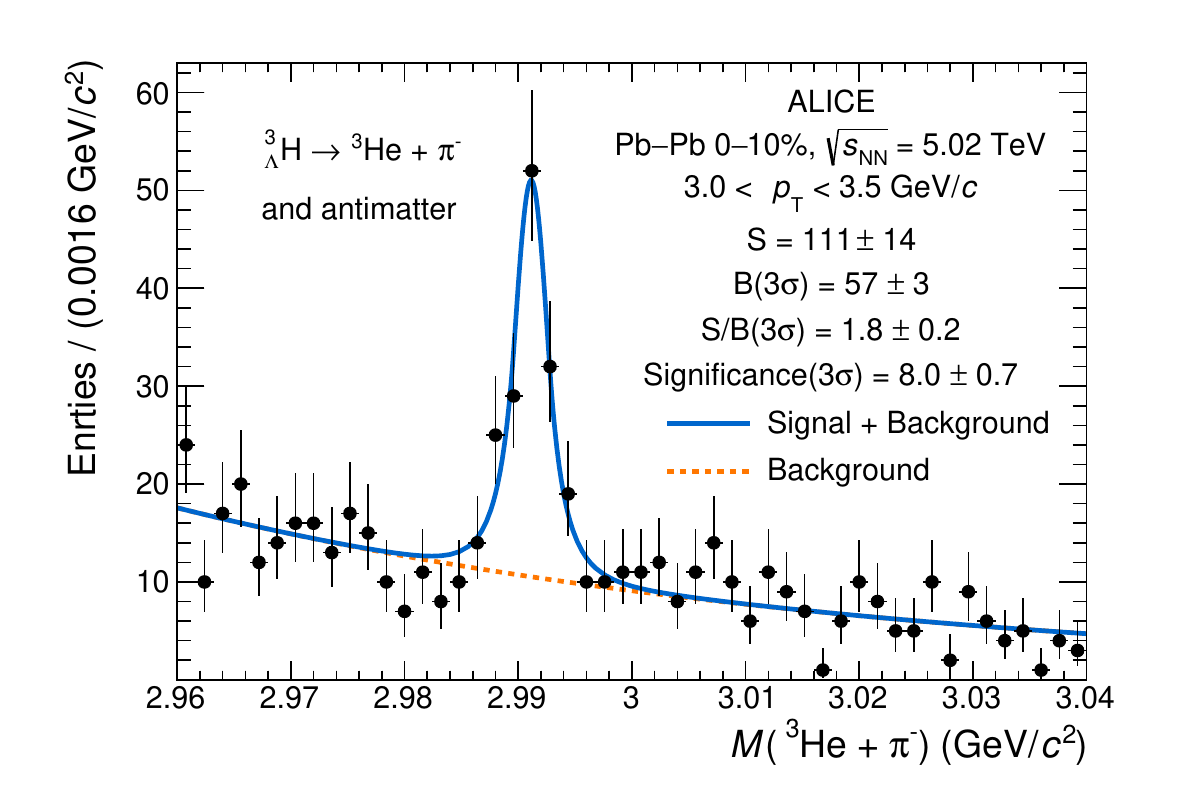}

    \caption{
    Invariant-mass distribution of selected \hyp candidates in the centrality class 0--10\% and \pt interval 3 $<$ \pt $<$ 3.5 \GeVc fitted with a
    function which is the sum of a double-sided Crystal Ball signal and an exponential background. Vertical lines represent the statistical uncertainties.
    }
    \label{fig:invfit}
\end{figure}

The main selection step consists of combining the candidate features in a gradient-boosted decision tree classifier (BDT)~\cite{10.1145/2939672.2939785, hipe4ml} that is employed for the \hyp signal selection.
The BDT is a supervised learning algorithm that determines how to differentiate between two or more classes, specifically signal and background in this context, by analysing sets of examples referred to as the training sets. In this analysis, the training sets consist of \hyp signal candidates obtained from the MC sample and background candidates from paired like-sign $\mathrm{^3He}$ and $\mathrm{\pi}$ tracks from the data.
For each \hyp candidate, the BDT combines a set of topological and single-track variables to produce a score, which serves to separate signal and background. 
The variables do not exhibit correlation with the \he-$\pi$ invariant mass and are: the decay length in the rest frame $ct$, the $\cos\theta_{\mathrm{P}}$, the PID information for \he, the DCA to the primary vertex for both \he and $\pi$ tracks, the DCA between \he and $\pi$, and the number of clusters of the \he track in the TPC.
Candidates with a BDT score exceeding a specified threshold are classified as signal. This threshold is determined to optimise the expected signal significance, taking into account the predicted production yield according to the SHM for the \hyp, as well as the background rate observed from like-sign \he and $\pi$ pairs.

The BDT training and testing steps and the \hyp yield computation have been performed in different centrality and \pt intervals independently.
The \hyp signal is extracted from the invariant-mass distribution after rejecting candidates whose score is lower than the BDT model threshold. An unbinned maximum-likelihood fit is performed to the \hyp invariant-mass distribution, employing a double-sided Crystal Ball (DSCB)~\cite{Verkerke:2003ir} function and an exponential function to model the signal and the background components of the spectrum, respectively. 
The parameters of the DSCB function are fixed using the MC sample leaving only the mean and the normalisation of the DSCB free. Figure~\ref{fig:invfit} shows an example of the invariant-mass fit in the most central collisions and in the \pt interval \mbox{$3 < \pt < 3.5$ \GeVc}.
The number of detected \hyp is obtained from the integral of the DSCB function. However, this number has to be corrected to account for the detection efficiency, the branching ratio (B.R.), and the absorption of the \hyp in the ALICE detector material. The detection efficiency comprises several components, including the acceptance of the ALICE detector, the absorption of daughter tracks, and the selection efficiency. 
It can be directly computed using the MC sample in different centrality and \pt intervals. The detection efficiency increases with the \hyp \pt and ranges from 10\% to 38\%. The branching ratio is set to 0.25 $\pm$ 0.023 according to Ref.~\cite{Kamada:1997rv}.
Finally, the absorption correction ($C_{\mathrm{abs}}$) is computed by simulating the interaction of \hyp with the ALICE detector material with \textsc{Geant4}, and evaluating the probability ($P_{\mathrm{abs}}$) of the \hyp to be absorbed before decaying. 
The absorption cross section employed for the simulation has been extracted from Ref.~\cite{Evlanov:1998py}, which accounts for the extension of the \hyp wave function, and amounts to about 1.5 times that of the \he.
The probability of \hyp to be absorbed is found to be increasing as a function of \pt due to the larger amount of material crossed by the high-momentum particle. The corresponding correction, computed as $C_{\mathrm{abs}}= 1-P_{\mathrm{abs}}$, ranges from 0.96 to 0.93.

The main systematic uncertainties on the measurement of the \hyp \pt spectra are due to the \hyp decay reconstruction, the BDT selection, and the invariant-mass fit in each \pt interval.
These systematic uncertainties are estimated as an envelope by employing a multi-trial approach.
In this approach, the following elements of the analysis are varied: two different weak-decay vertex-reconstruction methods, two different background datasets for the BDT training, two different fit functions for the \hyp signal (a Kernel Density Estimator~\cite{cranmer_kernel_2001} and the DSCB), three different fit functions for the combinatorial background modelling (polynomial of the first and second order, and an exponential function), and different BDT selection thresholds (giving rise to a $\pm 10\%$ variation of the efficiency around its nominal value). 
These elements are varied incoherently to populate a distribution of different yields that constitute our trial distribution in each \pt and centrality interval. The systematic uncertainty is then estimated as the standard deviation of the obtained distributions and varies from 8\% to 34\% in different intervals. 
Since the absorption cross section of \hyp is uncertain, an additional uncertainty due to the absorption is evaluated by changing the interaction cross section in the simulation to three times the \he inelastic cross section with the detector materials. This results in systematic uncertainties of 4\%--6\%, depending on the \pt interval.
Another source of the uncertainty originates from the uncertainty on the branching ratio, which amounts to 9\% and is common to all the \pt and centrality classes.

\section{Results and Discussions\label{sec:results}}

\begin{figure}[ht]
    \centering
    \includegraphics[width=0.99\textwidth]{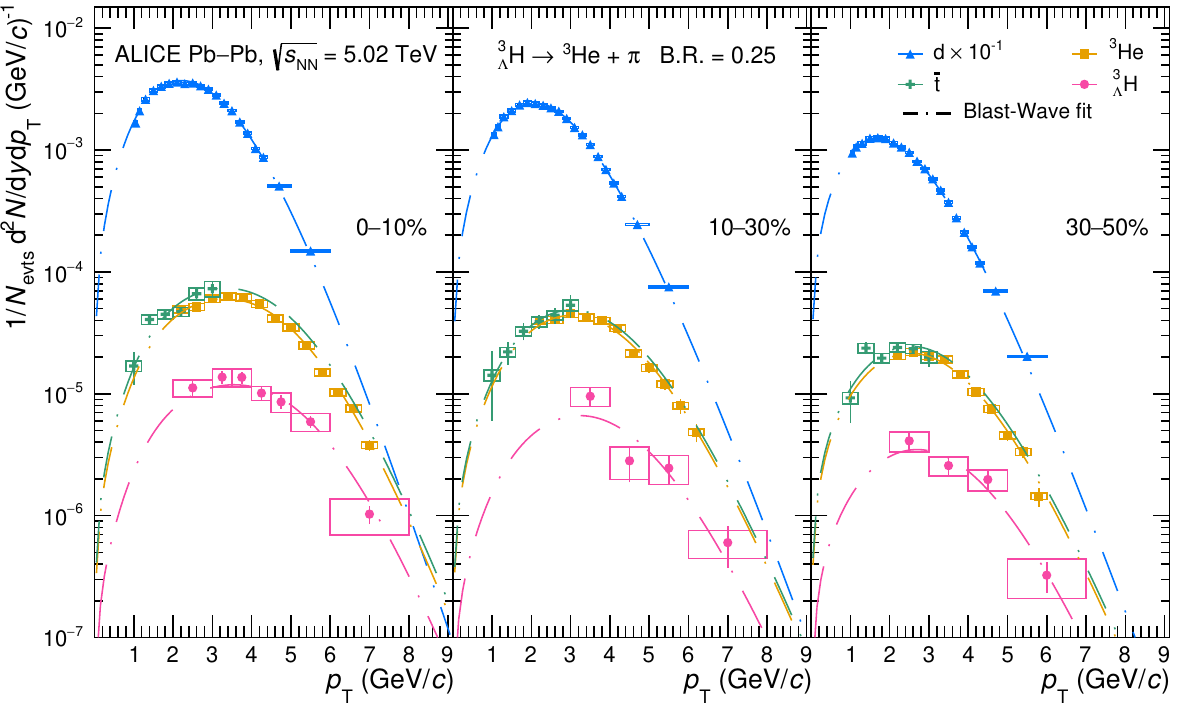}

    \caption{Deuteron (d), antitriton ($\mathrm{\overline{t}}$), \he~\cite{ALICE:2022veq}, and \hyp spectra measured in \PbPb collisions at \snn = 5.02 TeV. Each panel shows a centrality interval and different nuclei are reported with different colours. For \hyp, the average spectra between particles and antiparticles is employed.
    The width of the boxes refers to the range of the \pt intervals, while their height represents the systematic uncertainties. The vertical lines are the statistical uncertainties. The combined Blast-Wave fit parameters for d, $\mathrm{\overline{t}}$, \he, and \hyp are listed in Table~\ref{table:bwpara}.}
    \label{fig:bwfit}
\end{figure}

The \hyp \pt-differential spectra in different centrality intervals, defined as the average of particles and antiparticles, are shown in Fig.~\ref{fig:bwfit} along with the spectra of deuterons, antitritons, and \he in the corresponding centrality ranges, as measured in Ref.~\cite{ALICE:2022veq}. A simultaneous fit with the Blast-Wave model parameterisation~\cite{Schnedermann:1993ws} is done with all the shown spectra. The parameters of the combined Blast-Wave fit are listed in Table~\ref{table:bwpara}. Notably, the temperature and velocity in the Blast-Wave fit are compatible with those obtained by fitting only light nuclei (d, t, \he, \ensuremath{^4\mathrm{He}}\xspace)~\cite{ALICE:2023qyl}, suggesting a similar kinetic freeze-out surface for light-flavoured nuclei.

Individual Blast-Wave fits are employed to extrapolate the \hyp spectrum in the unmeasured \pt range and obtain the total yield ($\mathrm{d}N/\mathrm{d}y$) in the three centrality classes. The variation intervals of the fit parameters are restricted to an interval of $\pm 1\sigma$ around the parameters obtained from the combined fits shown in Fig.~\ref{fig:bwfit}. The resulting $\mathrm{d}N/\mathrm{d}y$ values are reported in Table~\ref{table:bwpara}. The fraction of extrapolated yield using the Blast-Wave fit is 13\%, 39\%, and 27\% for the 0--10\%, 10--30\%, and 30--50\% centrality classes, respectively. 
The total uncertainties on the integrated yields $\sigma(\mathrm{BW_{tot}})$ are computed by fitting the spectra considering both statistical and systematic uncertainties. To separate the statistical and systematic components of the uncertainty which originates from the Blast-Wave extrapolation, a fit considering only statistical uncertainty of the \hyp spectra is used to estimate the statistical component ($\sigma(\mathrm{BW_{stat}})$). The systematic uncertainty due to the fitting procedure is extracted using the relation $\sigma(\mathrm{BW_{syst}})=\sqrt{\sigma(\mathrm{BW_{tot}})^2 - \mathrm{\sigma(BW_{stat})^2}}$.
An additional source of systematic uncertainty to the measured yield arises from the unknown \pt distribution of \hyp.
As the fits performed with other functions, namely $m_{\mathrm{T}}$-exponential and Boltzmann, have not passed the Pearson $\chi^2$ test for the goodness of the fit and since the \pt distribution of \hyp should be very similar to \he according to the Blast-Wave model~\cite{Schnedermann:1993ws}, the related uncertainty is inherited from the previous \he production analysis~\cite{ALICE:2022veq} and corresponds to 1.8\%, 3.3\%, and 0.3\% of the measured integrated yield in 0--10\%, 10--30\%, 30--50\%, respectively.

\tabcolsep=0.13cm
\begin{table}[ht]
\centering
\caption{Parameters of the combined Blast-Wave fits and integrated (\hyp + \antihyp)/2 yields in different centrality intervals. The fits include deuteron, antitriton, \he, and \hyp as shown in Fig.~\ref{fig:bwfit}. The \hyp yields are obtained with an individual Blast-Wave fit whose parameters are restricted in 1$\sigma$ region of the parameters shown in the table. }
\label{table:bwpara}
\begin{tabular}{cccccc}
\toprule
Centrality & $\langle\beta_\mathrm{T}\rangle$ & T~(\GeV) & \it{n} & $\chi^2$ / ndf & $\mathrm{d}N/\mathrm{d}y \times 10^{-5}$ (B.R. = 0.25)\\
\midrule
0--10\% &   0.694   $\pm$   0.003  & 0.103   $\pm$   0.005 & 0.498 $\pm$ 0.009 & 43.4 / 39 & 4.83   $\pm$   0.23(stat) $\pm$ 0.57(syst) \\
10--30\% & 0.666   $\pm$   0.003  & 0.132   $\pm$   0.008 & 0.507 $\pm$ 0.012 & 19.1 / 34 & 2.62   $\pm$   0.25(stat) $\pm$ 0.40(syst) \\
30--50\% & 0.598   $\pm$   0.005 &  0.152   $\pm$   0.010 & 0.660 $\pm$ 0.022 & 21.8 / 33 & 1.27   $\pm$   0.10(stat) $\pm$ 0.14(syst)\\
\bottomrule
\end{tabular}
\vspace{0.1in}
\end{table}

The production yield of \hyp provides a powerful tool to investigate the mechanism of nuclear production in relativistic hadronic collisions~\cite{ALICE:2022sco}. The measured \hyp yield in all the centrality intervals is directly compared with the predictions from the grand-canonical SHM~\cite{Vovchenko:2019pjl} with T=155 MeV. The predicted \hyp $\mathrm{d}N/\mathrm{d}y$ ($9.5\times 10^{-5}$) is approximately two times higher than the measured one in central \PbPb collisions, suggesting that the SHM, as applied, may lack certain elements necessary for accurately describing the generation of this weakly bound state. 

\begin{figure}[t]
    \centering
    \includegraphics[width=0.7\textwidth]{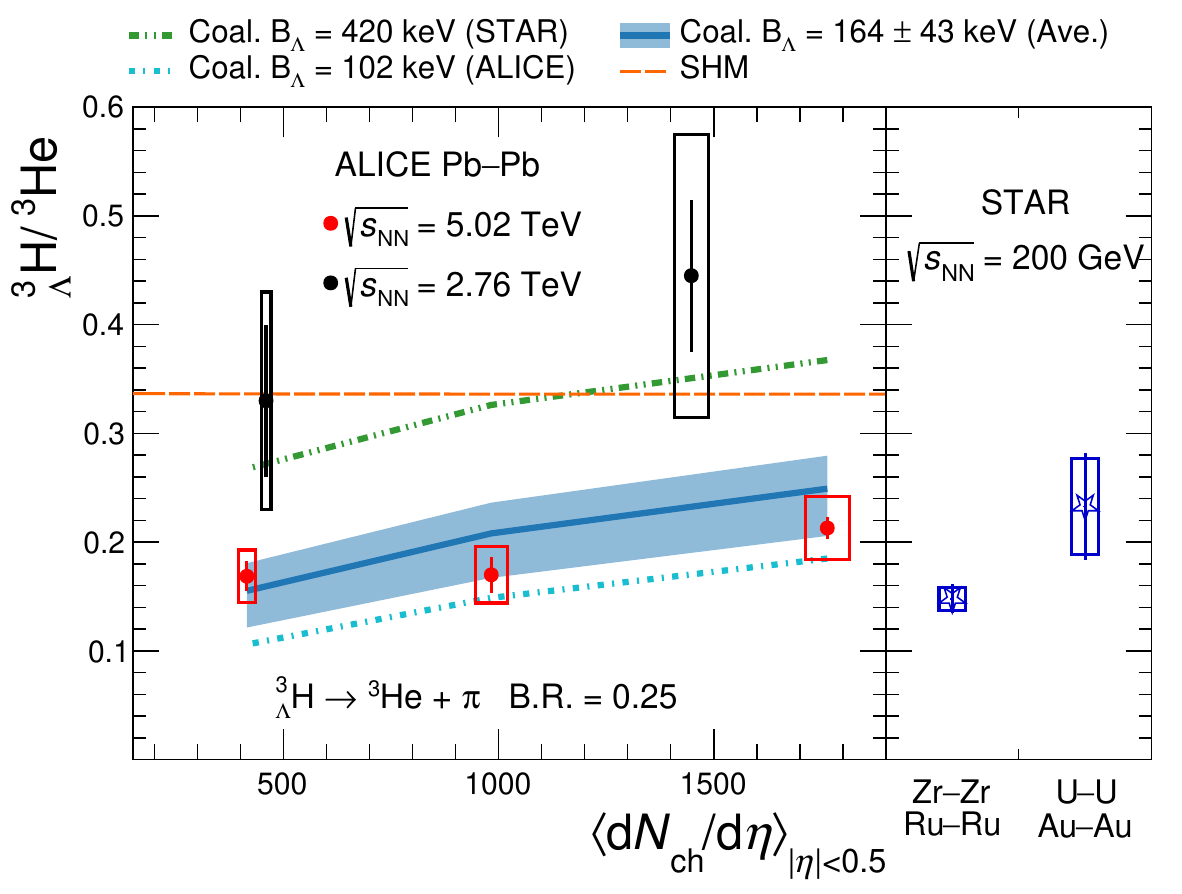}

    \caption{Yield ratio of \hyp to \he together with theoretical predictions as a function of multiplicity. In the left panel, the results of this analysis are compared with the ALICE measurement at \snn = 2.76 TeV ~\cite{ALICE:2015oer}. Vertical lines are used for the statistical uncertainties and the height of the boxes for the systematic ones. The width of the boxes refers to the uncertainties of multiplicity. For each centrality interval the $\langle \mathrm{d}N_\mathrm{ch}/\mathrm{d}\eta\rangle$ is taken from Ref.~\cite{ALICE:2015juo} and the \he yield from Ref.~\cite{ALICE:2022veq}. The dense orange dashed line represents the expectation of SHM, while the other three sets of lines stand for coalescence model with different $B_\Lambda$ hypotheses. 
    The coalescence prediction with world average $B_\Lambda$ is displayed with a 1$\sigma$ uncertainty as the filled area, both lines and shadowed areas are linear interpolations of the available model calculations~\cite{Liu:2024ygk}. In the right panel, the results of recent STAR measurement are shown for comparison~\cite{STAR:2023fbc}. }
    \label{fig:ratiomul}
\end{figure}

On the other hand, while the SHM is able to calculate directly the absolute yields of hadrons, in the coalescence model only the yield ratios among particles can be computed without any further knowledge of the momentum spectra of the nucleons. To compare with both the models, SHM and coalescence, the $\hyp/\he$ ratio is considered. The \he production spectra from Ref.~\cite{ALICE:2022veq} are not corrected for the feed-down from \hyp. However, this effect is estimated to be at most 1.7\%, and a systematic uncertainty of the same magnitude is included in the ratio to account for it.
The produced charged-particle multiplicities per unit of pseudorapidity $\langle\mathrm{d}N_{\mathrm{ch}}/\mathrm{d}\eta\rangle$, which are related to the centrality of the collision and can reflect the size of the fireball, may have effects on the ratios of particle yields and need to be taken into account. In this case, the grand-canonical SHM predicts no multiplicity dependence. 
A deviation from the grand-canonical ensemble prediction is present in the canonical SHM only for $\langle\mathrm{d}N_{\mathrm{ch}}/\mathrm{d}\eta\rangle_{|\eta|<0.5}<100$~\cite{Vovchenko:2018fiy}, which corresponds to peripheral collisions comparatively and is not covered by the presented results.
The coalescence model formulation here considered~\cite{Liu:2024ygk} uses the Wigner function formalism and a parametrisation of the wave function of the \hyp that depends on its $\Lambda$ separation energy, and it was already successfully applied to predict the production of other light nuclei in heavy-ion collisions~\cite{Zhao:2021dka, Sun:2022rjh}.
Three different coalescence predictions are shown by setting $B_\Lambda$ to the latest ALICE and STAR experimental values and to the world average of the $B_\Lambda$ measurements~\cite{ALICE:2022sco, STAR:2021orx, HypernuclearDataBase}. The uncertainty due to the $B_\Lambda$ value is asymmetric as the \hyp wave function width is not a linear function of $B_\Lambda$. 
Furthermore, the coalescence model predicts a suppression of the $\hyp/\he$ ratio with decreasing multiplicities due to the interplay between the smaller hadron emission source in peripheral Pb--Pb collisions and the wide wave function of the \hyp.
In Fig.~\ref{fig:ratiomul}, the model predictions are compared with the measured $\hyp/\he$ yield ratio as a function of the average charged-particle multiplicity of the analysed centrality intervals. 
The measured $\hyp/\he$ yield ratios are well described by the coalescence prediction using the current world average of $B_{\Lambda}$ (solid blue line). In addition, the recent STAR measurements~\cite{STAR:2023fbc} suggest a similar suppression of the $\hyp/\he$ ratio going from large (Au--Au and U--U) to smaller (Zr--Zr and Ru--Ru) collision systems. For larger values of $B_{\Lambda}$, the predictions from the coalescence model approach that of the SHM, but they are not compatible with the measured ratios. 
The results from a prior ALICE analysis~\cite{ALICE:2015oer} in \PbPb collisions at $\sqrt{s_{\mathrm{NN}}}$ = 2.76 \TeV are included in Fig.~\ref{fig:ratiomul} for comparison. The results presented in this Letter remain consistent with the previous study within a 2$\sigma$ confidence interval, but uncertainties are twice as small as those obtained from the smaller data sample at lower collision energy.

\begin{figure}[ht]
    \centering
    \includegraphics[width=\textwidth]{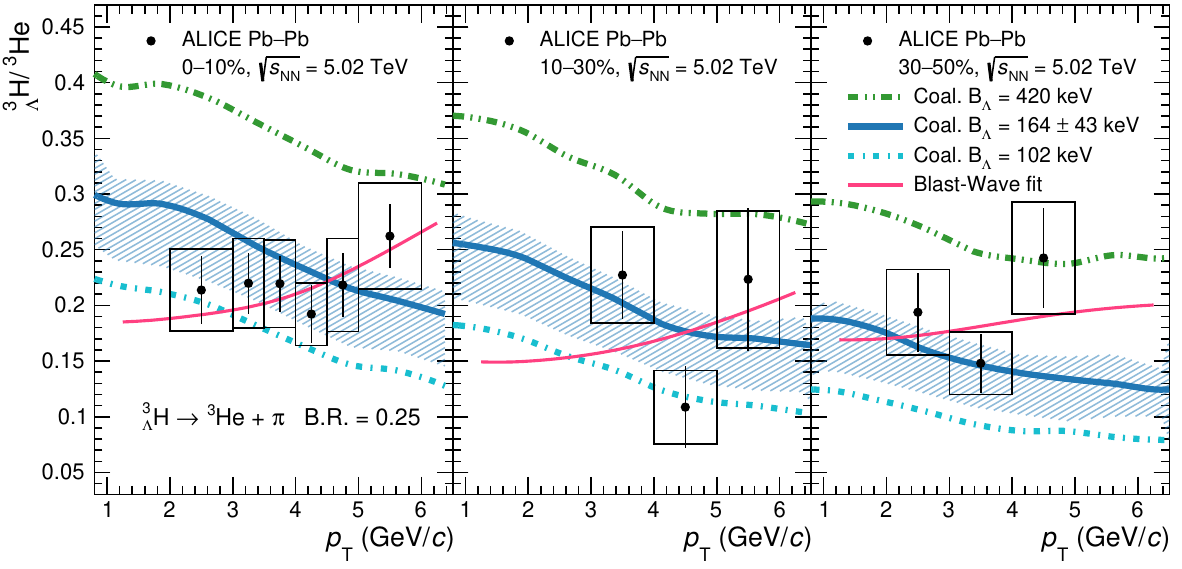}

    \caption{The $\hyp/\he$ yield ratio together with theoretical predictions in different centrality intervals as a function of \pt. The vertical lines and the height of the boxes represent the statistical and systematic uncertainties, respectively. The width of the boxes displays the range of the \pt intervals. The pink solid lines are calculated as the ratios of the Blast-Wave fit functions for \he and \hyp. The other three curves stand for the predictions of the coalescence model with different $B_\Lambda$ hypotheses. 
    The coalescence prediction with world average $B_\Lambda$ is displayed with a 1$\sigma$ uncertainty band~\cite{Liu:2024ygk}.}
    \label{fig:ratiopt}
\end{figure}

In Fig.~\ref{fig:ratiopt}, the $\hyp/\he$ yield ratio as a function of \pt in the three different centrality intervals is shown. While the SHM does not infer the expected \pt shape of the particles, the coalescence mechanism implies a decrease of the $\hyp/\he$ ratio for increasing \pt. Conversely, in a simple hydrodynamic picture like the Blast-Wave model, the radial flow boosts the high \pt production of heavier particles such as the \hyp. Consequently, within the Blast-Wave framework, the $\hyp/\he$ ratio is expected to increase as a function of \pt. 
However, the current experimental uncertainties do not allow for a definitive conclusion on the trend of the \hyp / \he ratio as a function of \pt.

\section{Conclusions}
This Letter presents the first \pt-differential measurement of the \hyp production in \PbPb collisions at \fivenn. The \pt distribution of \hyp is well described by a simultaneous Blast-Wave fit with other light nuclei. The fit temperature and velocity profiles are compatible with those found for other light-flavoured nuclei, hinting to a common kinetic freeze-out surface for hypernuclei and ordinary nuclei produced in \PbPb collisions. 
The \pt-integrated yields show a significant discrepancy with respect to the predictions of the SHM tuned to fit the other light-flavoured particles.
Furthermore, the yield ratio $\hyp/\he$ was calculated to test the predictions of the coalescence model, that is able to describe the measured ratios when assuming the correct binding energy of the \hyp.
Finally, both the coalescence and the Blast-Wave models describe the experimental data for the $\hyp/\he$ ratios as a function of \pt. However, their predictions have a different \pt trend. Presently, the large experimental uncertainties preclude a definitive interpretation; however, the forthcoming data from LHC Run 3 will allow the ALICE Collaboration to measure this quantity with unprecedented precision.

\newenvironment{acknowledgement}{\relax}{\relax}
\begin{acknowledgement}
\section*{Acknowledgements}

The ALICE Collaboration would like to thank all its engineers and technicians for their invaluable contributions to the construction of the experiment and the CERN accelerator teams for the outstanding performance of the LHC complex.
The ALICE Collaboration gratefully acknowledges the resources and support provided by all Grid centres and the Worldwide LHC Computing Grid (WLCG) collaboration.
The ALICE Collaboration acknowledges the following funding agencies for their support in building and running the ALICE detector:
A. I. Alikhanyan National Science Laboratory (Yerevan Physics Institute) Foundation (ANSL), State Committee of Science and World Federation of Scientists (WFS), Armenia;
Austrian Academy of Sciences, Austrian Science Fund (FWF): [M 2467-N36] and Nationalstiftung f\"{u}r Forschung, Technologie und Entwicklung, Austria;
Ministry of Communications and High Technologies, National Nuclear Research Center, Azerbaijan;
Conselho Nacional de Desenvolvimento Cient\'{\i}fico e Tecnol\'{o}gico (CNPq), Financiadora de Estudos e Projetos (Finep), Funda\c{c}\~{a}o de Amparo \`{a} Pesquisa do Estado de S\~{a}o Paulo (FAPESP) and Universidade Federal do Rio Grande do Sul (UFRGS), Brazil;
Bulgarian Ministry of Education and Science, within the National Roadmap for Research Infrastructures 2020-2027 (object CERN), Bulgaria;
Ministry of Education of China (MOEC) , Ministry of Science \& Technology of China (MSTC) and National Natural Science Foundation of China (NSFC), China;
Ministry of Science and Education and Croatian Science Foundation, Croatia;
Centro de Aplicaciones Tecnol\'{o}gicas y Desarrollo Nuclear (CEADEN), Cubaenerg\'{\i}a, Cuba;
Ministry of Education, Youth and Sports of the Czech Republic, Czech Republic;
The Danish Council for Independent Research | Natural Sciences, the VILLUM FONDEN and Danish National Research Foundation (DNRF), Denmark;
Helsinki Institute of Physics (HIP), Finland;
Commissariat \`{a} l'Energie Atomique (CEA) and Institut National de Physique Nucl\'{e}aire et de Physique des Particules (IN2P3) and Centre National de la Recherche Scientifique (CNRS), France;
Bundesministerium f\"{u}r Bildung und Forschung (BMBF) and GSI Helmholtzzentrum f\"{u}r Schwerionenforschung GmbH, Germany;
General Secretariat for Research and Technology, Ministry of Education, Research and Religions, Greece;
National Research, Development and Innovation Office, Hungary;
Department of Atomic Energy Government of India (DAE), Department of Science and Technology, Government of India (DST), University Grants Commission, Government of India (UGC) and Council of Scientific and Industrial Research (CSIR), India;
National Research and Innovation Agency - BRIN, Indonesia;
Istituto Nazionale di Fisica Nucleare (INFN), Italy;
Japanese Ministry of Education, Culture, Sports, Science and Technology (MEXT) and Japan Society for the Promotion of Science (JSPS) KAKENHI, Japan;
Consejo Nacional de Ciencia (CONACYT) y Tecnolog\'{i}a, through Fondo de Cooperaci\'{o}n Internacional en Ciencia y Tecnolog\'{i}a (FONCICYT) and Direcci\'{o}n General de Asuntos del Personal Academico (DGAPA), Mexico;
Nederlandse Organisatie voor Wetenschappelijk Onderzoek (NWO), Netherlands;
The Research Council of Norway, Norway;
Pontificia Universidad Cat\'{o}lica del Per\'{u}, Peru;
Ministry of Science and Higher Education, National Science Centre and WUT ID-UB, Poland;
Korea Institute of Science and Technology Information and National Research Foundation of Korea (NRF), Republic of Korea;
Ministry of Education and Scientific Research, Institute of Atomic Physics, Ministry of Research and Innovation and Institute of Atomic Physics and Universitatea Nationala de Stiinta si Tehnologie Politehnica Bucuresti, Romania;
Ministry of Education, Science, Research and Sport of the Slovak Republic, Slovakia;
National Research Foundation of South Africa, South Africa;
Swedish Research Council (VR) and Knut \& Alice Wallenberg Foundation (KAW), Sweden;
European Organization for Nuclear Research, Switzerland;
Suranaree University of Technology (SUT), National Science and Technology Development Agency (NSTDA) and National Science, Research and Innovation Fund (NSRF via PMU-B B05F650021), Thailand;
Turkish Energy, Nuclear and Mineral Research Agency (TENMAK), Turkey;
National Academy of  Sciences of Ukraine, Ukraine;
Science and Technology Facilities Council (STFC), United Kingdom;
National Science Foundation of the United States of America (NSF) and United States Department of Energy, Office of Nuclear Physics (DOE NP), United States of America.
In addition, individual groups or members have received support from:
Czech Science Foundation (grant no. 23-07499S), Czech Republic;
European Research Council (grant no. 950692), European Union;
ICSC - Centro Nazionale di Ricerca in High Performance Computing, Big Data and Quantum Computing, European Union - NextGenerationEU;
Academy of Finland (Center of Excellence in Quark Matter) (grant nos. 346327, 346328), Finland.

\end{acknowledgement}

\bibliographystyle{utphys}   
\bibliography{bibliography}

\newpage
\appendix

%
%

\section{The ALICE Collaboration}
\label{app:collab}
\begin{flushleft} 
\small

S.~Acharya\,\orcidlink{0000-0002-9213-5329}\,$^{\rm 127}$, 
D.~Adamov\'{a}\,\orcidlink{0000-0002-0504-7428}\,$^{\rm 86}$, 
A.~Agarwal$^{\rm 135}$, 
G.~Aglieri Rinella\,\orcidlink{0000-0002-9611-3696}\,$^{\rm 32}$, 
L.~Aglietta\,\orcidlink{0009-0003-0763-6802}\,$^{\rm 24}$, 
M.~Agnello\,\orcidlink{0000-0002-0760-5075}\,$^{\rm 29}$, 
N.~Agrawal\,\orcidlink{0000-0003-0348-9836}\,$^{\rm 25}$, 
Z.~Ahammed\,\orcidlink{0000-0001-5241-7412}\,$^{\rm 135}$, 
S.~Ahmad\,\orcidlink{0000-0003-0497-5705}\,$^{\rm 15}$, 
S.U.~Ahn\,\orcidlink{0000-0001-8847-489X}\,$^{\rm 71}$
% ...

\section*{Affiliation Notes}

$^{\rm I}$ Deceased\\
$^{\rm II}$ Also at: Max-Planck-Institut fur Physik, Munich, Germany\\
$^{\rm III}$ Also at: Italian National Agency for New Technologies, Energy and Sustainable Economic Development (ENEA), Bologna, Italy\\
$^{\rm IV}$ Also at: Dipartimento DET del Politecnico di Torino, Turin, Italy\\
$^{\rm V}$ Also at: Yildiz Technical University, Istanbul, T\"{u}rkiye\\
$^{\rm VI}$ Also at: Department of Applied Physics, Aligarh Muslim University, Aligarh, India\\
$^{\rm VII}$ Also at: Institute of Theoretical Physics, University of Wroclaw, Poland\\
$^{\rm VIII}$ Also at: An institution covered by a cooperation agreement with CERN\\

\section*{Collaboration Institutes}

$^{1}$ A.I. Alikhanyan National Science Laboratory (Yerevan Physics Institute) Foundation, Yerevan, Armenia\\
$^{2}$ AGH University of Krakow, Cracow, Poland\\
$^{3}$ Bogolyubov Institute for Theoretical Physics, National Academy of Sciences of Ukraine, Kiev, Ukraine\\
$^{4}$ Bose Institute, Department of Physics  and Centre for Astroparticle Physics and Space Science (CAPSS), Kolkata, India\\
$^{5}$ California Polytechnic State University, San Luis Obispo, California, United States\\
$^{6}$ Central China Normal University, Wuhan, China\\
$^{7}$ Centro de Aplicaciones Tecnol\'{o}gicas y Desarrollo Nuclear (CEADEN), Havana, Cuba\\
$^{8}$ Centro de Investigaci\'{o}n y de Estudios Avanzados (CINVESTAV), Mexico City and M\'{e}rida, Mexico\\
$^{9}$ Chicago State University, Chicago, Illinois, United States\\
$^{10}$ China Institute of Atomic Energy, Beijing, China\\
$^{11}$ China University of Geosciences, Wuhan, China\\
$^{12}$ Chungbuk National University, Cheongju, Republic of Korea\\
$^{13}$ Comenius University Bratislava, Faculty of Mathematics, Physics and Informatics, Bratislava, Slovak Republic\\
$^{14}$ Creighton University, Omaha, Nebraska, United States\\
$^{15}$ Department of Physics, Aligarh Muslim University, Aligarh, India\\
$^{16}$ Department of Physics, Pusan National University, Pusan, Republic of Korea\\
$^{17}$ Department of Physics, Sejong University, Seoul, Republic of Korea\\
$^{18}$ Department of Physics, University of California, Berkeley, California, United States\\
$^{19}$ Department of Physics, University of Oslo, Oslo, Norway\\
$^{20}$ Department of Physics and Technology, University of Bergen, Bergen, Norway\\
$^{21}$ Dipartimento di Fisica, Universit\`{a} di Pavia, Pavia, Italy\\
$^{22}$ Dipartimento di Fisica dell'Universit\`{a} and Sezione INFN, Cagliari, Italy\\
$^{23}$ Dipartimento di Fisica dell'Universit\`{a} and Sezione INFN, Trieste, Italy\\
$^{24}$ Dipartimento di Fisica dell'Universit\`{a} and Sezione INFN, Turin, Italy\\
$^{25}$ Dipartimento di Fisica e Astronomia dell'Universit\`{a} and Sezione INFN, Bologna, Italy\\
$^{26}$ Dipartimento di Fisica e Astronomia dell'Universit\`{a} and Sezione INFN, Catania, Italy\\
$^{27}$ Dipartimento di Fisica e Astronomia dell'Universit\`{a} and Sezione INFN, Padova, Italy\\
$^{28}$ Dipartimento di Fisica `E.R.~Caianiello' dell'Universit\`{a} and Gruppo Collegato INFN, Salerno, Italy\\
$^{29}$ Dipartimento DISAT del Politecnico and Sezione INFN, Turin, Italy\\
$^{30}$ Dipartimento di Scienze MIFT, Universit\`{a} di Messina, Messina, Italy\\
$^{31}$ Dipartimento Interateneo di Fisica `M.~Merlin' and Sezione INFN, Bari, Italy\\
$^{32}$ European Organization for Nuclear Research (CERN), Geneva, Switzerland\\
$^{33}$ Faculty of Electrical Engineering, Mechanical Engineering and Naval Architecture, University of Split, Split, Croatia\\
$^{34}$ Faculty of Engineering and Science, Western Norway University of Applied Sciences, Bergen, Norway\\
$^{35}$ Faculty of Nuclear Sciences and Physical Engineering, Czech Technical University in Prague, Prague, Czech Republic\\
$^{36}$ Faculty of Physics, Sofia University, Sofia, Bulgaria\\
$^{37}$ Faculty of Science, P.J.~\v{S}af\'{a}rik University, Ko\v{s}ice, Slovak Republic\\
$^{38}$ Frankfurt Institute for Advanced Studies, Johann Wolfgang Goethe-Universit\"{a}t Frankfurt, Frankfurt, Germany\\
$^{39}$ Fudan University, Shanghai, China\\
$^{40}$ Gangneung-Wonju National University, Gangneung, Republic of Korea\\
$^{41}$ Gauhati University, Department of Physics, Guwahati, India\\
$^{42}$ Helmholtz-Institut f\"{u}r Strahlen- und Kernphysik, Rheinische Friedrich-Wilhelms-Universit\"{a}t Bonn, Bonn, Germany\\
$^{43}$ Helsinki Institute of Physics (HIP), Helsinki, Finland\\
$^{44}$ High Energy Physics Group,  Universidad Aut\'{o}noma de Puebla, Puebla, Mexico\\
$^{45}$ Horia Hulubei National Institute of Physics and Nuclear Engineering, Bucharest, Romania\\
$^{46}$ HUN-REN Wigner Research Centre for Physics, Budapest, Hungary\\
$^{47}$ Indian Institute of Technology Bombay (IIT), Mumbai, India\\
$^{48}$ Indian Institute of Technology Indore, Indore, India\\
$^{49}$ INFN, Laboratori Nazionali di Frascati, Frascati, Italy\\
$^{50}$ INFN, Sezione di Bari, Bari, Italy\\
$^{51}$ INFN, Sezione di Bologna, Bologna, Italy\\
$^{52}$ INFN, Sezione di Cagliari, Cagliari, Italy\\
$^{53}$ INFN, Sezione di Catania, Catania, Italy\\
$^{54}$ INFN, Sezione di Padova, Padova, Italy\\
$^{55}$ INFN, Sezione di Pavia, Pavia, Italy\\
$^{56}$ INFN, Sezione di Torino, Turin, Italy\\
$^{57}$ INFN, Sezione di Trieste, Trieste, Italy\\
$^{58}$ Inha University, Incheon, Republic of Korea\\
$^{59}$ Institute for Gravitational and Subatomic Physics (GRASP), Utrecht University/Nikhef, Utrecht, Netherlands\\
$^{60}$ Institute of Experimental Physics, Slovak Academy of Sciences, Ko\v{s}ice, Slovak Republic\\
$^{61}$ Institute of Physics, Homi Bhabha National Institute, Bhubaneswar, India\\
$^{62}$ Institute of Physics of the Czech Academy of Sciences, Prague, Czech Republic\\
$^{63}$ Institute of Space Science (ISS), Bucharest, Romania\\
$^{64}$ Institut f\"{u}r Kernphysik, Johann Wolfgang Goethe-Universit\"{a}t Frankfurt, Frankfurt, Germany\\
$^{65}$ Instituto de Ciencias Nucleares, Universidad Nacional Aut\'{o}noma de M\'{e}xico, Mexico City, Mexico\\
$^{66}$ Instituto de F\'{i}sica, Universidade Federal do Rio Grande do Sul (UFRGS), Porto Alegre, Brazil\\
$^{67}$ Instituto de F\'{\i}sica, Universidad Nacional Aut\'{o}noma de M\'{e}xico, Mexico City, Mexico\\
$^{68}$ iThemba LABS, National Research Foundation, Somerset West, South Africa\\
$^{69}$ Jeonbuk National University, Jeonju, Republic of Korea\\
$^{70}$ Johann-Wolfgang-Goethe Universit\"{a}t Frankfurt Institut f\"{u}r Informatik, Fachbereich Informatik und Mathematik, Frankfurt, Germany\\
$^{71}$ Korea Institute of Science and Technology Information, Daejeon, Republic of Korea\\
$^{72}$ KTO Karatay University, Konya, Turkey\\
$^{73}$ Laboratoire de Physique Subatomique et de Cosmologie, Universit\'{e} Grenoble-Alpes, CNRS-IN2P3, Grenoble, France\\
$^{74}$ Lawrence Berkeley National Laboratory, Berkeley, California, United States\\
$^{75}$ Lund University Department of Physics, Division of Particle Physics, Lund, Sweden\\
$^{76}$ Nagasaki Institute of Applied Science, Nagasaki, Japan\\
$^{77}$ Nara Women{'}s University (NWU), Nara, Japan\\
$^{78}$ National and Kapodistrian University of Athens, School of Science, Department of Physics , Athens, Greece\\
$^{79}$ National Centre for Nuclear Research, Warsaw, Poland\\
$^{80}$ National Institute of Science Education and Research, Homi Bhabha National Institute, Jatni, India\\
$^{81}$ National Nuclear Research Center, Baku, Azerbaijan\\
$^{82}$ National Research and Innovation Agency - BRIN, Jakarta, Indonesia\\
$^{83}$ Niels Bohr Institute, University of Copenhagen, Copenhagen, Denmark\\
$^{84}$ Nikhef, National institute for subatomic physics, Amsterdam, Netherlands\\
$^{85}$ Nuclear Physics Group, STFC Daresbury Laboratory, Daresbury, United Kingdom\\
$^{86}$ Nuclear Physics Institute of the Czech Academy of Sciences, Husinec-\v{R}e\v{z}, Czech Republic\\
$^{87}$ Oak Ridge National Laboratory, Oak Ridge, Tennessee, United States\\
$^{88}$ Ohio State University, Columbus, Ohio, United States\\
$^{89}$ Physics department, Faculty of science, University of Zagreb, Zagreb, Croatia\\
$^{90}$ Physics Department, Panjab University, Chandigarh, India\\
$^{91}$ Physics Department, University of Jammu, Jammu, India\\
$^{92}$ Physics Program and International Institute for Sustainability with Knotted Chiral Meta Matter (SKCM2), Hiroshima University, Hiroshima, Japan\\
$^{93}$ Physikalisches Institut, Eberhard-Karls-Universit\"{a}t T\"{u}bingen, T\"{u}bingen, Germany\\
$^{94}$ Physikalisches Institut, Ruprecht-Karls-Universit\"{a}t Heidelberg, Heidelberg, Germany\\
$^{95}$ Physik Department, Technische Universit\"{a}t M\"{u}nchen, Munich, Germany\\
$^{96}$ Politecnico di Bari and Sezione INFN, Bari, Italy\\
$^{97}$ Research Division and ExtreMe Matter Institute EMMI, GSI Helmholtzzentrum f\"ur Schwerionenforschung GmbH, Darmstadt, Germany\\
$^{98}$ Saga University, Saga, Japan\\
$^{99}$ Saha Institute of Nuclear Physics, Homi Bhabha National Institute, Kolkata, India\\
$^{100}$ School of Physics and Astronomy, University of Birmingham, Birmingham, United Kingdom\\
$^{101}$ Secci\'{o}n F\'{\i}sica, Departamento de Ciencias, Pontificia Universidad Cat\'{o}lica del Per\'{u}, Lima, Peru\\
$^{102}$ Stefan Meyer Institut f\"{u}r Subatomare Physik (SMI), Vienna, Austria\\
$^{103}$ SUBATECH, IMT Atlantique, Nantes Universit\'{e}, CNRS-IN2P3, Nantes, France\\
$^{104}$ Sungkyunkwan University, Suwon City, Republic of Korea\\
$^{105}$ Suranaree University of Technology, Nakhon Ratchasima, Thailand\\
$^{106}$ Technical University of Ko\v{s}ice, Ko\v{s}ice, Slovak Republic\\
$^{107}$ The Henryk Niewodniczanski Institute of Nuclear Physics, Polish Academy of Sciences, Cracow, Poland\\
$^{108}$ The University of Texas at Austin, Austin, Texas, United States\\
$^{109}$ Universidad Aut\'{o}noma de Sinaloa, Culiac\'{a}n, Mexico\\
$^{110}$ Universidade de S\~{a}o Paulo (USP), S\~{a}o Paulo, Brazil\\
$^{111}$ Universidade Estadual de Campinas (UNICAMP), Campinas, Brazil\\
$^{112}$ Universidade Federal do ABC, Santo Andre, Brazil\\
$^{113}$ Universitatea Nationala de Stiinta si Tehnologie Politehnica Bucuresti, Bucharest, Romania\\
$^{114}$ University of Cape Town, Cape Town, South Africa\\
$^{115}$ University of Derby, Derby, United Kingdom\\
$^{116}$ University of Houston, Houston, Texas, United States\\
$^{117}$ University of Jyv\"{a}skyl\"{a}, Jyv\"{a}skyl\"{a}, Finland\\
$^{118}$ University of Kansas, Lawrence, Kansas, United States\\
$^{119}$ University of Liverpool, Liverpool, United Kingdom\\
$^{120}$ University of Science and Technology of China, Hefei, China\\
$^{121}$ University of South-Eastern Norway, Kongsberg, Norway\\
$^{122}$ University of Tennessee, Knoxville, Tennessee, United States\\
$^{123}$ University of the Witwatersrand, Johannesburg, South Africa\\
$^{124}$ University of Tokyo, Tokyo, Japan\\
$^{125}$ University of Tsukuba, Tsukuba, Japan\\
$^{126}$ Universit\"{a}t M\"{u}nster, Institut f\"{u}r Kernphysik, M\"{u}nster, Germany\\
$^{127}$ Universit\'{e} Clermont Auvergne, CNRS/IN2P3, LPC, Clermont-Ferrand, France\\
$^{128}$ Universit\'{e} de Lyon, CNRS/IN2P3, Institut de Physique des 2 Infinis de Lyon, Lyon, France\\
$^{129}$ Universit\'{e} de Strasbourg, CNRS, IPHC UMR 7178, F-67000 Strasbourg, France, Strasbourg, France\\
$^{130}$ Universit\'{e} Paris-Saclay, Centre d'Etudes de Saclay (CEA), IRFU, D\'{e}partment de Physique Nucl\'{e}aire (DPhN), Saclay, France\\
$^{131}$ Universit\'{e}  Paris-Saclay, CNRS/IN2P3, IJCLab, Orsay, France\\
$^{132}$ Universit\`{a} degli Studi di Foggia, Foggia, Italy\\
$^{133}$ Universit\`{a} del Piemonte Orientale, Vercelli, Italy\\
$^{134}$ Universit\`{a} di Brescia, Brescia, Italy\\
$^{135}$ Variable Energy Cyclotron Centre, Homi Bhabha National Institute, Kolkata, India\\
$^{136}$ Warsaw University of Technology, Warsaw, Poland\\
$^{137}$ Wayne State University, Detroit, Michigan, United States\\
$^{138}$ Yale University, New Haven, Connecticut, United States\\
$^{139}$ Yonsei University, Seoul, Republic of Korea\\
$^{140}$  Zentrum  f\"{u}r Technologie und Transfer (ZTT), Worms, Germany\\
$^{141}$ Affiliated with an institute covered by a cooperation agreement with CERN\\
$^{142}$ Affiliated with an international laboratory covered by a cooperation agreement with CERN.\\

\end{flushleft} 

\end{document}